\documentstyle[12pt]{article}

\newcommand{\nc}{\newcommand}
\def\frac#1#2{{\textstyle {#1 \over #2}}}

\nc{\beq}{\begin{equation}}
\nc{\eeq}{\end{equation}}
\nc{\beqa}{\begin{eqnarray}}
\nc{\eeqa}{\end{eqnarray}}
\nc{\lsim}{\begin{array}{c}\,\sim\vspace{-21pt}\\< \end{array}}
\nc{\gsim}{\begin{array}{c}\sim\vspace{-21pt}\\> \end{array}}

\def\&{and}

\def\DS {D\!\!\!\!/}

\def\nc#1#2#3{           {\it Nuovo Cim.  }{\bf #1}, #2 (19#3)}
\def\np#1#2#3{           {\it Nucl. Phys. }{\bf #1}, #2 (19#3)}
\def\pl#1#2#3{           {\it Phys. Lett. }{\bf #1}, #2 (19#3)}
\def\pr#1#2#3{           {\it Phys. Rev. }{\bf #1}, #2 (19#3)}
\def\prep#1#2#3{         {\it Phys. Rep. }{\bf #1}, #2 (19#3)}
\def\prl#1#2#3{          {\it Phys. Rev. Lett. }{\bf #1}, #2 (19#3)}

\begin{document}

\begin{titlepage}

\begin{center}
January 1996      \hfill       YCTP-P3-95\\
\vskip .5 in
{\large \bf TOPOLOGICAL CHARGE AND $U(1)_A$  SYMMETRY\\
IN THE HIGH TEMPERATURE PHASE OF QCD}
\vskip .3 in
{    {\bf
Nick Evans\footnote{nick@zen.physics.yale.edu},
Stephen D.H. Hsu\footnote{hsu@hsunext.physics.yale.edu}
    and Myckola Schwetz\footnote{ms@genesis2.physics.yale.edu}
}
}

   \vskip 0.3 cm
   {\it Sloane Physics Laboratory,
        Yale University,
        New Haven, CT 06511}\\
\end{center}

\vskip .5 in
\begin{abstract}
\noindent
We discuss the global symmetries of the high temperature
phase of  QCD  with $N_f$ massless quarks. We show that the
$U(N_f) \times U(N_f)$ symmetries are only violated by
operators of dimension $\geq 3 N_f$. For $N_f > 2$ this
implies that the thermal two-point  correlation functions of the
$\eta'$ and $\pi^a$'s are identical. We discuss the implications of this
for the chiral phase transition at finite temperature.

\end{abstract}
\end{titlepage}

\renewcommand{\thepage}{\arabic{page}}
\setcounter{page}{1}


Recently, Cohen \cite{TC} has used QCD inequalties
to argue that the high temperature phase of massless QCD
with $N_f$ flavors of quarks is
effectively symmetric under a global $U(N_f) \times U(N_f)$ symmetry
rather than just $SU(N_f) \times SU(N_f)$.
Cohen examines the
two-point correlation functions of different operators
such as the $\eta'$ ($\bar{q} \gamma_5 q$)
and
the $\pi^a$ ($\bar{q} \tau^a \gamma_5 q$)
which are in the same
$U(N_f) \times U(N_f)$ multiplet but {\it not} in the
same $SU(N_f) \times SU(N_f)$ multiplet.
He argues that in the massless (or chiral)
limit: $m_q \rightarrow 0$, the difference between
the respective two-point correlation functions
approaches zero.

In this letter we show that the $U(N_f) \times U(N_f)$ symmetry is not
completely restored in the high temperature phase
although its breaking can only be manifested in operators of
dimension $\geq 3N_f$. Thus the $U(N_f) \times U(N_f)$ symmetry is
only restored for the two-point correlators when $N_f > 2$.
We concentrate on  the relationship between (spontaneous)
chiral symmetry breaking, the axial anomaly and
topological charge and clarify some subtle points in the argument of \cite{TC}.

It is well-known that the solution of the $U(1)_A$
problem \cite{U1} requires
\vskip .2 in
\noindent (A) the axial anomaly
relation  ($m_q  = 0$)
\beq
\label{anomaly}
\partial_{\mu} J_{\mu}^5 ~=~ \frac{g^2 N_f}{16 \pi^2} tr[ F \tilde{F} ],
\eeq
and
\vskip .2 in
\noindent (B)
the presence of gauge configurations with non-zero
topological charge
\beq
\nu \equiv \frac{1}{16 \pi^2} \int d^4x~ tr[ F \tilde{F} ].
\eeq
\vskip .2 in

We can understand this by examining the behavior of
correlators under $U(1)_A$ transformations.
Let\footnote{We assume throughout this paper
a non-perturbative
regulator which conserves chiral symmetry
and is controlled by a UV scale $\Lambda$.
See, e.g., \cite{reg}.}

\beq
\label{O}
\langle {\cal O} \rangle ~=~ \frac{1}{Z} \int D[A] ~e^{-S_{YM}}~
 \int D \psi D \bar{\psi} e^{- \int \bar{\psi} (i \DS - m_q) \psi}~
{\cal O}
\eeq
where ${\cal O}$ is an operator built out of $\psi, \bar{\psi}$
and gauge fields. Now consider ${\cal O}$ after an
axial transformation:
\beq
\label{Oa}
{\cal O} \rightarrow {\cal O}_{\alpha} ~=~ {\cal O} ~(\bar{\psi}', \psi')
\eeq
where
\beqa
\label{rotate}
\psi &\rightarrow& \psi' ~=~  e^{i \alpha \gamma_5} \psi    \\
\bar{\psi} &\rightarrow& \bar{\psi} ' ~=~
\bar{\psi} e^{i \alpha \gamma_5}.
\eeqa
The change in the expectation of ${\cal O}$ is given by
\beqa
\label{CWI}
\delta_{\alpha} \langle {\cal O} \rangle &~=~&
\langle {\cal O}_{\alpha} \rangle -
\langle {\cal O} \rangle  \nonumber \\
&~=~& \frac{1}{Z} \int D[A] ~e^{-S_{YM}}~
 \int D \psi D \bar{\psi} e^{- \int \bar{\psi} (i \DS - m_q) \psi}~
\left[ {\cal O}_{\alpha} - {\cal O} \right].
\eeqa
The integral in (\ref{CWI}) can be evaluated by a change
of variables: $\psi \rightarrow \psi'$, $\bar{\psi} \rightarrow \bar{\psi}'$.
The only subtlety is that the Jacobian for the change of variables
induces the anomaly factor in the measure of the functional
integral \cite{F}. In the chiral limit, this yields
\beq
\label{CWI1}
\delta_{\alpha} \langle {\cal O} \rangle ~=~
\langle ~[e^{2 i \alpha \nu} - 1] ~{\cal O} \rangle
{}~=~ \sum_{\nu \neq 0}~ \langle ~[e^{2 i \alpha \nu} - 1]
{}~{\cal O} \rangle_{\nu},
\eeq
where $~\langle ~~~~~ \rangle_{\nu}~$ denotes the expectation
value taken in the sector of the gauge field configuration
space with topological charge $\nu$.
Thus we see that the physical effect of the $U(1)_A$ anomaly
is only manifested in sectors of the functional integral with
non-zero topological charge.

The operator relation in (\ref{anomaly}),
which we used to derive (\ref{CWI1}) follows from the
ultraviolet (UV) behavior of the theory and
is therefore unaffected by temperature. One can easily
see this by repeating the Fujikawa derivation \cite{F}
with boundary conditions in the Euclidean time direction
which are appropriate for $T \neq 0$. As long as the
UV regularization scale is kept large compared to $T$
the same result for the anomalous variation of the functional
measure is obtained.
However, we will demonstrate below that above the
temperature at which chiral symmetry is restored, the contributions
of gauge
configurations with $\nu \neq 0$ to correlators of quark operators of
dimension $< 3N_f$ are suppressed
(they are in fact a set of `measure zero' in the functional measure)
and hence
the axial anomaly has no effect on the $U(1)_A$ Ward
identities for these operators.
For these operators we will show that the right hand side
of (\ref{CWI1}) is zero when there is no spontaneous
chiral symmetry breaking (i.e. in the high temperature
phase of QCD) and the current quark masses $m_q$ are taken to
zero.

The presence of massless quarks is known to suppress
topological fluctuations.
The partition function for QCD is
\beqa
\label{Z}
Z ~&=&~ \int D[A] ~e^{-S_{YM}}~ Det[ \DS - m_q ]~
e^{i \theta \int F \tilde{F}}    \nonumber \\
&\equiv&~ \int [d\mu_A] ~e^{i \theta \int F \tilde{F}}
\eeqa
where $Det[ \DS - m_q ] ~=~ \prod_n ~ (i \lambda_n - m_q)$ and
the $\lambda_n$ are eigenvalues of the Euclidean
Dirac equation: $\DS \psi_n = i \lambda_n \psi_n$.
We can break $Z$ into contributions from sectors of
different winding number $\nu$:
\beq
\label{Zwind}
Z ~=~
\sum_{\nu = - \infty}^{\nu = + \infty}~ Z_{\nu}~ e^{i \theta \nu}.
\eeq
An index theorem \cite{U1} tells us that there must exist
a minimum number of zero mode solutions when $\nu \neq 0$:
$\nu ~=~ n_+ - n_-$, where $n_{\pm}$ is the number
of right(left)-handed solutions. This implies that in the
$\nu \neq 0$ sectors $Z_{\nu}$ must vanish at least
as fast as  $m_q^{\vert \nu \vert N_f}$
in the chiral limit. (In this paper we will assume
exact $SU(N)_V$ isospin symmetry and take the quark mass
matrix to be proportional to the identity matrix: $M = m_q {\cal I}$.)
This seems to imply that only the $\nu = 0$ sector contributes
to $Z$ when massless quarks are present.

When chiral symmetry breaking is involved, this argument is too
naive. This is fortunate since we believe that in QCD,
which exhibits chiral symmetry breaking, the $U(1)_A$ problem is
indeed solved by the combination of (A) and (B).
Heuristically, one might guess that the
`dynamical mass' acquired by the quarks plays some role
in determining whether topological fluctuations are suppressed.
This issue has been addressed systematically by
Leutwyler and Smilga \cite{LS}. They find that fluctuations
in topological charge are controlled by the parameter
$X = \Sigma~ V m_q$, where
\beq
\label{Sig}
\Sigma = \lim_{m \rightarrow 0} \lim_{V \rightarrow \infty}
\frac{1}{V} \int d^4x~\langle \bar{q} q (x)\rangle
\eeq
is the chiral symmetry breaking order parameter, and $V$ is
the volume of the system. At large $X$ topological fluctuations
are allowed, whereas as $X \rightarrow 0$ they are suppressed.
The order of limits in taking $V \rightarrow \infty$ and $m_q \rightarrow 0$
is important to the discussion here. If one takes $m_q \rightarrow 0$
with $V$ fixed one does not recover the phase of QCD in which
chiral symmetries are spontaneously broken. This is because symmetries
cannot break spontaneously at finite volume, and hence for fixed $V$
a non-zero $m_q$ is required to bias the vacuum energy and keep
the system in the broken vacuum. It is only at $X >> 1$ that finite-volume
effects are small and we recover the low-energy vacuum of QCD.
However in this limit topological fluctuations are unsuppressed
even though $m_q$ goes to zero. This is essentially a consequence of chiral
symmetry breaking.

It is believed that in the high temperature phase of QCD
chiral symmetry is restored by thermal effects. In other
words, at sufficiently high temperature,
the minimum of the free energy is at $\Sigma = 0$.
We can study the theory at finite temperature by
imposing (anti-) periodic boundary conditions in Euclidean
time on the (fermion) boson fields appearing in the
functional integral. If the period $\beta = 1/T$ is taken to
be sufficiently small we will recover the high temperature phase.
In this phase the subtlety associated with the order of limits
$m_q \rightarrow 0$, $V \rightarrow \infty$ is
no longer important\footnote{For the order of limits to
be important,  some physical quantities would have to
depend on
parameters such as, e.g., $m_q L$,
where $L$ is the size of our box. However, this is highly
implausible in the high-T (disordered) phase as there are
neither long-range order nor long-range correlations
in the heat bath.}
 as we are not trying to recover a phase
with spontaneous symmetry breaking. Instead, we can
take $m_q \rightarrow 0$ {\it before} taking the $V \rightarrow \infty$
(or the UV cutoff $\Lambda \rightarrow \infty$). This has
important consequences, as it allows the naive scaling arguments
in $m_q$ to be used. It also allows us to make
our arguments rigorous, as all quantities are finite  while
$V$ and $\Lambda$ are kept finite.
For example, in the chirally restored phase
$Z_{\nu}$ does indeed vanish like
$m_q^{\vert \nu \vert N_f}$ in the chiral limit.
In the absence of dynamical chiral symmetry breaking
massless quarks suppress the contributions to $Z$ from
sectors of non-trivial topology.

We next consider the contributions from sectors with
$\nu \neq 0$
to arbitrary correlators
in the high-temperature phase.
By performing the fermionic part of the functional integral,
the right hand side of (\ref{CWI1}) can
be written as a sum over permutations of functional integrals over
the gauge field measure
with integrands that are
products of traces of  propagators, gamma matrices and flavor matrices
(times possibly some functions of the gauge field, which we suppress):
\beq
\label{ad}
\delta_\alpha \langle {\cal O} \rangle = \sum_{\nu \neq 0}
\sum_{\rm perms} \int [d\mu_{A}]_{~\nu}~ [e^{i \alpha \nu} -1 ]~
\prod~ tr \left[ S_A^1(x_i, x_j) \Gamma_1  S_A^2(x_k, x_l)
\Gamma_2... \right] ,
 \eeq
where $\Gamma$ is any combination of gamma and flavor
matrices.

In each $\nu$ sector the measure, $\int [d\mu_A]_{~\nu}$,
approaches zero as $m_q^{n_0}$ where $n_0$ is the number of
zero modes and hence $n_0 \geq \vert \nu \vert N_f$.
The propagators may be written in terms of their spectral decomposition
\beq
S_A (x,y) = \sum_k {\psi^\dagger_k(x) \psi_k(y) \over \lambda_k -im_q}
\eeq
where it is important to remember that Fermi statistics (imposed by the
integration
over Grassmanian variables) forbids any two propagators in the product of sums
from sharing an eigenvalue $\lambda_k$.
Thus the integrands in (\ref{ad}) diverge at most
as $(1/m_q)^{n}$ where the operator $\cal O$ is of dimension $3n$
but where the power of divergence must always be $\leq n_0$.
The functional integral will never diverge as a result of the
$m_q \rightarrow 0$ limit since the propagators in ${\cal O}$ may
at most `soak up' all $n_0$ zero modes \footnote{This is a loophole in
the argument of
\cite{TC} where it is argued that if the contribution to
an an n-point correlator from a particular $\nu$ sector
is non-zero the (n+2)-point quark correlator
must diverge.  If the n-point correlator is non-zero all zero modes
in that sector have been soaked up by the propagators and the introduction
of further propagators will not generate any further inverse powers of $m_q$.
This loophole is problematic for the results of \cite{TC} when $N_f = 2$.}.
However, we observe that
correlators of all operators of dimension $< 3N_f$ receive no
contributions from gauge configurations with $\nu \neq 0$
in the $m_q \rightarrow 0$ limit.
The operators whose expectation values in $\nu \neq 0$ sectors
are non-vanishing when $m_q \rightarrow 0$ are precisely the
so-called 'tHooft operators \cite{tH} induced by instanton
processes at weak coupling.

The two-point correlators are of special interest since they determine
the number of (nearly) massless modes present at and
above the high temperature phase transition,
which is believed to be either second or weakly
first order. For example, consider
the two-point functions for the $\pi$ and $\eta'$
at finite temperature:
\beqa
\label{tp}
\langle \eta'(x) \eta'(0) \rangle ~&=&~
\langle ~\bar{\psi}_i \gamma_5 \psi_i~ (x)
{}~ \bar{\psi}_j \gamma_5 \psi_j ~ (0) \rangle \\
\langle \pi^a (x) \pi^a (0) \rangle ~&=&~
\langle~ \bar{\psi} \tau^a \gamma_5 \psi~(x)
{}~ \bar{\psi} \tau^a \gamma_5 \psi ~(0) ~\rangle
\eeqa
Writing these correlators in terms of
exact quark propagators $S_A (x,y)$,
one finds two types of contributions:
a disconnected contribution
\beq
\label{dc}
\frac{1}{Z}\int d \mu_A ~tr[\Gamma S_A(x,x)] Tr[\Gamma S_A (0,0)]
\eeq
and a connected part
\beq
\label{c}
\frac{1}{Z} \int d \mu_A ~ tr[S_A (x,0) \Gamma S_A(x,0) \Gamma].
\eeq
Here $\Gamma = \gamma_5$ for the $\eta'$ and
$\Gamma = \tau^a \gamma_5$ for the $\pi^a$.
The connected parts (\ref{c}) are identical since
$[ \tau^a, S_A ] = 0$.
For the pion, the disconnected part is zero since
$tr[ \tau^a] = 0$. Any $\eta'$-$\pi^a$ splitting is the result of
(\ref{dc}) for the $\eta'$.

The measure in a given $\nu$ sector goes to zero at least as fast as
$m_q^{\vert \nu \vert N_f}$
as $m_q \rightarrow 0$  whilst the two propagators may
soak up only two of the zero modes. Thus for $N_f > 2$ the difference between
the correlators vanishes as $m \rightarrow 0$ and the $\eta'$ is
degenerate with the massless pions. A similar result holds for the entire
$U(N_f) \times U(N_f)$ multiplet which consists of the $\pi$,$\delta$,
$\sigma$ and $\eta'$ resonances.

When $N_f \leq 2$, however, there are potentially contributions to the
correlator difference from the $\nu = \pm1$ sector.
There is good reason to believe that these contributions
are non-zero since  one-instanton effects which contribute to
(\ref{dc}) were found in the
weak coupling regime by
't~Hooft \cite{tH}.  His calculations in the weak coupling approximation  are
relevant at temperatures $T \gg \Lambda_{QCD}$,
where the effective coupling constant is small and
non-perturbative effects can be studied in the
semiclassical approximation (see also \cite{KY}).
When $N_f = 2$ instanton effects contribute directly to
(\ref{dc}), leading to an $\eta'$--$\pi^a$ mass splitting.
It is possible in principle that near the chiral phase transition
$T \simeq T_c$, where the dilute instanton analysis is not
completely reliable, that some other effects (such as $I$-$\bar{I}$
pair formation \cite{YAK}) lead to the suppression of this
splitting, but it is implausible that
(\ref{dc}) vanishes completely.

In the case of $N_f  = 1$ there are
potentially contributions to the chiral condensate
\beq \Sigma = {1 \over Z} \int d\mu_A ~ tr[S_A(x,x)]  \eeq
from the $\nu= \pm1$ sector. These can be seen to be non-zero
even at high-temperature \cite{KY}, and
combining this with the low temperature analysis of  Leutwyler and
Smilga \cite{LS}, one reaches the conclusion that the $U(1)_A$
violating
chiral condensate stays non-zero for all temperatures with
$U(1)_V$ being the only unbroken global symmetry.

We end with a summary of our conclusions for
different values of $N_f$, and the corresponding implications
for the chiral phase transition.

\begin{itemize}

{\item   $N_f = 1$:
There is no chiral phase transition,
$~\langle \bar{\psi} \psi \rangle ~$  remains non-zero at
high temperatures, and
the dynamically
generated quark mass decreases
smoothly to zero as $ T \rightarrow \infty $.}

{\item  $N_f = 2$:
The  $SU(2) \times SU(2)$ global symmetry is restored in
the high temperature phase.
The $\eta'$--$\pi^a$  splitting is non-zero, but
decreases smoothly  to zero with temperature
as determined by the instanton density:
${m_{\eta'}}^2 \sim ({\Lambda \over T})^k$.
At $T \simeq T_c$ the $\eta'$ is massive and the symmetry
remains $SU(2) \times SU(2)$.
Renormalization group calculations \cite{PW} (based on the
$\varepsilon$-expansion)
in an effective linear sigma model  with this symmetry indicate that
the phase transition is second order and
QCD lattice simulations \cite{KR} appear to be in agreement with
this prediction.} It is worth mentioning that if the
$\eta'$-$\pi^a$ splitting is non-zero but {\it small}
at $T \simeq T_c$, the effective model for the system as
$T \rightarrow T_c$ from above
may have an approximate $U(2) \times U(2)$ symmetry.
Studies of $U(2) \times U(2)$ dynamics \cite{PW}
generically lead to a fluctuation-induced first order transition,
so it is possible that the chiral phase transition occurs
through this instability before reaching the second order
fixed point at length scales much larger than the $\eta'$
correlation length. For an overview of these and related issues,
see \cite{YAK}.

{\item  $N_f \geq 3$ (assuming asymptotic freedom):
The $U(N_f) \times U(N_f)$ global symmetry is effectively restored
(up to high-dimension operators which are probably irrelevant
in the IR limit)
in the high temperature phase. The $\eta'$
becomes degenerate with the $\pi^a$'s
 for $T \geq T_c $. The effective models of this chiral phase
transition should
incorporate a $U(N_f) \times U(N_f)$ global symmetry.
The analysis of  the appropriate linear
sigma model  without determinantal interactions \cite{PW, ASS}
suggests that  the phase transitions is first  order with its
strength increasing with $N_f$.}

{\item Finally, it is amusing to note that in
QCD theories with $N_f$  just below ${11 \over 2} N_c$,
such that the theory has a perturbative infra-red fixed point,
there is no dynamical chiral symmetry breaking and hence our
results also apply there independent of the temperature.
In such theories the $U(N_f) \times U(N_f)$ symmetry applies
to all correlators up to some high dimension, where there are some
exponentially small $U(1)_A$ violating interactions in the form
of high-dimension instanton operators ('tHooft vertices).
Strictly speaking there
are no mesons here since we are at weak coupling.}

\end{itemize}

\vskip .5 in
\centerline{\bf Acknowledgements}
\vskip 0.1in

The authors would like to thank T. Cohen for useful
comments. This work was supported under 

DOE contract DE-AC02-ERU3075.

\end{document}